\documentclass[runningheads]{llncs}
\usepackage{graphicx}
\usepackage{amsfonts}
\usepackage{amsmath}

\usepackage{subcaption}
\usepackage{graphicx}
\graphicspath{{figures/}}
\usepackage{float}
\usepackage{array}
\usepackage{color}
\usepackage{booktabs}

\usepackage[square, comma, sort&compress]{natbib}

\usepackage{hyperref}
\hypersetup{
    colorlinks=true,
    linkcolor=red,
    citecolor=green,
    filecolor=blue,
    urlcolor=magenta,
    linktocpage=true
}

\usepackage[dvipsnames]{xcolor} 

\makeatletter
\newcommand{\printfnsymbol}[1]{%
  \textsuperscript{\@fnsymbol{#1}}%
}
\makeatother

\newcommand{\bag}{\mathcal{X}} 
\newcommand{\instance}{x} 
\newcommand{\ssp}[1]{^{(#1)}} 
\newcommand{\DimEncode}{d_{\text{enc}}} 

\newcommand{\AttnVarFcn}[2]{\text{Var}\left(#1; #2\right)} 

\begin{document}









\title{Incorporating intratumoral heterogeneity into weakly-supervised deep learning models via variance pooling}

\titlerunning{Variance pooling for intratumoral heterogeneity}


%
\author{Iain Carmichael\thanks{These authors contributed equally} \and 
Andrew H. Song\printfnsymbol{1} \and 
Richard J. Chen \and
Drew F.K. Williamson \and
Tiffany Y. Chen \and
Faisal Mahmood}

\authorrunning{I. Carmichael, A.H. Song et al.}
\institute{Department of Pathology, Brigham and Women’s Hospital, Harvard Medical School, Boston, MA\\
Department of Pathology, Massachusetts General Hospital, Harvard Medical School, Boston, MA\\
\email{\{icarmichael, asong, rchen14, dwilliamson, tchen25, faisalmahmood\}@bwh.harvard.edu}}

\maketitle              

\begin{abstract}
Supervised learning tasks such as cancer survival prediction from gigapixel whole slide images (WSIs) are a critical challenge in computational pathology that requires modeling complex features of the tumor microenvironment.
These learning tasks are often solved with deep multi-instance learning (MIL) models that do not explicitly capture intratumoral heterogeneity.
We develop a novel variance pooling architecture that enables a MIL model to incorporate intratumoral heterogeneity into its predictions.
Two interpretability tools based on ``representative patches" are illustrated to probe the biological signals captured by these models.
An empirical study with 4,479 gigapixel WSIs from the Cancer Genome Atlas shows that adding variance pooling onto MIL frameworks improves  survival prediction performance for five cancer types.

\end{abstract}

\section{Introduction} \label{s:intro}

Cancer diagnosis and prognosis using \textit{hematoxylin and eosin} (H\&E) stained \textit{whole slide images} (WSIs) are central tasks in computational pathology. These supervised learning problems are made difficult by the size and complexity of WSIs; these images can be gigapixels in scale and involve localizing subtle visual signals that expert pathologists are trained for years to identify. In recent years, weakly supervised deep learning algorithms have made significant progress in addressing some of these challenges \citep{ coudray2018classification, couture2018image, campanella2019clinical, courtiol2019deep, wulczyn2021interpretable, lu2021data}.

Cancer cell morphology can vary greatly within a single tumor and it is increasingly recognized that \textit{intratumoral heterogeneity} (ITH) plays an important role in prognostication for certain cancer types \citep{andor2016pan,marusyk2020intratumor}, though its exact role is still being elucidated.
Furthermore, ITH complicates prognostication based on WSIs by increasing the range of morphologies that correspond to a particular prognosis.
For example, pathologists often encounter \textit{clear cell renal cell carcinoma} displaying significant heterogeneity within the same tumor, with some areas demonstrating classic clear cell morphology and others displaying morphologies bordering on diagnoses such as \textit{chromophobe renal cell carcinoma} or \textit{clear cell tubulopapillary renal cell carcinoma}.

Existing supervised algorithms that scale to WSIs do not generally incorporate histological ITH.
These algorithms typically approach a supervised learning task such as survival prediction through a \textit{multiple instance learning} (MIL) framework \citep{ilse18attention}.
First, a WSI is represented as a \textit{bag} of many smaller image patches.
Next, neural network features are extracted from each patch.
Finally, these patch features are aggregated for prediction.
The standard MIL aggregation approaches capture only simple statistics (e.g. mean or max) and neglect higher-order information (e.g. variance), which can capture heterogeneity.

To this end, we develop a novel MIL aggregation component that scales to gigapixel WSIs and explicitly incorporates ITH into its predictions.
We accomplish this through a \textit{variance pooling} operation that quantifies ITH as the variance along a collection of low rank projections of patch features (Section \ref{ss:attn_var_pool}).
Moving beyond pre-specified measures of heterogeneity (e.g. the variance of tumor cell morphology), this architecture uses the data to determine how to measure heterogeneity (e.g. the projection vectors, along which the variance of the features are measured, are learned.) 
This operation can be seamlessly incorporated into existing MIL architectures, as we now demonstrate.

We demonstrate empirically in Section \ref{s:exper} that including this operation improves baseline MIL models' performance on survival prediction tasks across five cancer types from The Cancer Genome Atlas (TCGA).
Through attention and a novel ``variance projection contrast" visualization, we provide interpretable insights into what biological signals these architectures utilize to make the predictions in Section \ref{s:interpret}.
Code: \url{https://github.com/mahmoodlab/VARPOOL}.

\subsection{Related Work}\label{ss:lit_review}
\noindent\textbf{Survival Analysis using WSIs:}
Survival analysis in histopathology has been a longstanding problem in medical imaging.
Early work required manual identification of regions of interest \citep{gurcan2009histopathological}.
Other methods make use of hand-crafted tumor cell features (e.g. shape) to predict patient survival \citep{lu2018nuclear}.
Most recent survival methods use  MIL algorithms that are directly trained on WSIs without additional annotations \citep{zhu2017wsisa, mobadersany2018predicting, wulczyn2021interpretable, zhao2020predicting}. 
Some of these methods use \textit{graph convolutional networks} (GCNs), which learn more context-aware features \citep{ zhao2020predicting, chen2021whole}.

\noindent\textbf{Intratumoral Heterogeneity:} Much of the previous work on ITH uses genomic data to identify and quantify subclones of tumor cells that share mutations \citep{raynaud2018pan}. 
There is growing interest in using machine learning to quantify the histologic ITH that pathologists observe on a daily basis \citep{faust2020unsupervised}.

\section{Incorporating heterogeneity with variance pooling}\label{s:method}

This section presents the novel variance pooling operation for weakly-supervised MIL tasks.
In this setting, we observe a set of bags -- an unordered collection of vectors (instances) -- that we use to predict some bag level response.
In computational pathology, each patient has a WSI which is used to predict the response (e.g. class label or survival outcome).
Each patent's WSI\footnote{When a patient has multiple WSIs, the patches are unioned across the WSIs.} (the bag) is broken up into many small image patches and the instances are either the raw image patches or patch features extracted by a neural network.

We first define some notation. 
Suppose we have $N$ patients.
For the $i$th patient, we observe a response $y\ssp{i}$ and a bag of instances (image patch features) $\bag\ssp{i}=\{\instance\ssp{i}_1, \ldots, \instance\ssp{i}_{n\ssp{i}}\}$, where each $\instance\ssp{i}_j \in \mathbb{R}^d$  and $n\ssp{i}$ is the size of the bag.
For classification tasks $y\ssp{i}$ is a patient-level class label; for survival tasks $y\ssp{i} = (t\ssp{i}, c\ssp{i})$ where $t\ssp{i} \in \mathbb{R}_+$ is the observed survival time and $c\ssp{i} \in \{0, 1\}$ indicates whether or not this time was censored.
We want to learn a mapping $f(\bag\ssp{i}) = z\ssp{i}$ that outputs a patient-level prediction minimizing a supervised loss function. 

In the following two sections, we  present the variance pool architecture in the context of the attention mean pooling model of \cite{ilse18attention}. 
We emphasize variance pooling can be easily incorporated into other MIL architectures.
For notational simplicity we generally drop the patient superscript $i$.  

\vspace{-1em}
\begin{figure}[!ht]
    \centering
    \includegraphics[width=\linewidth]{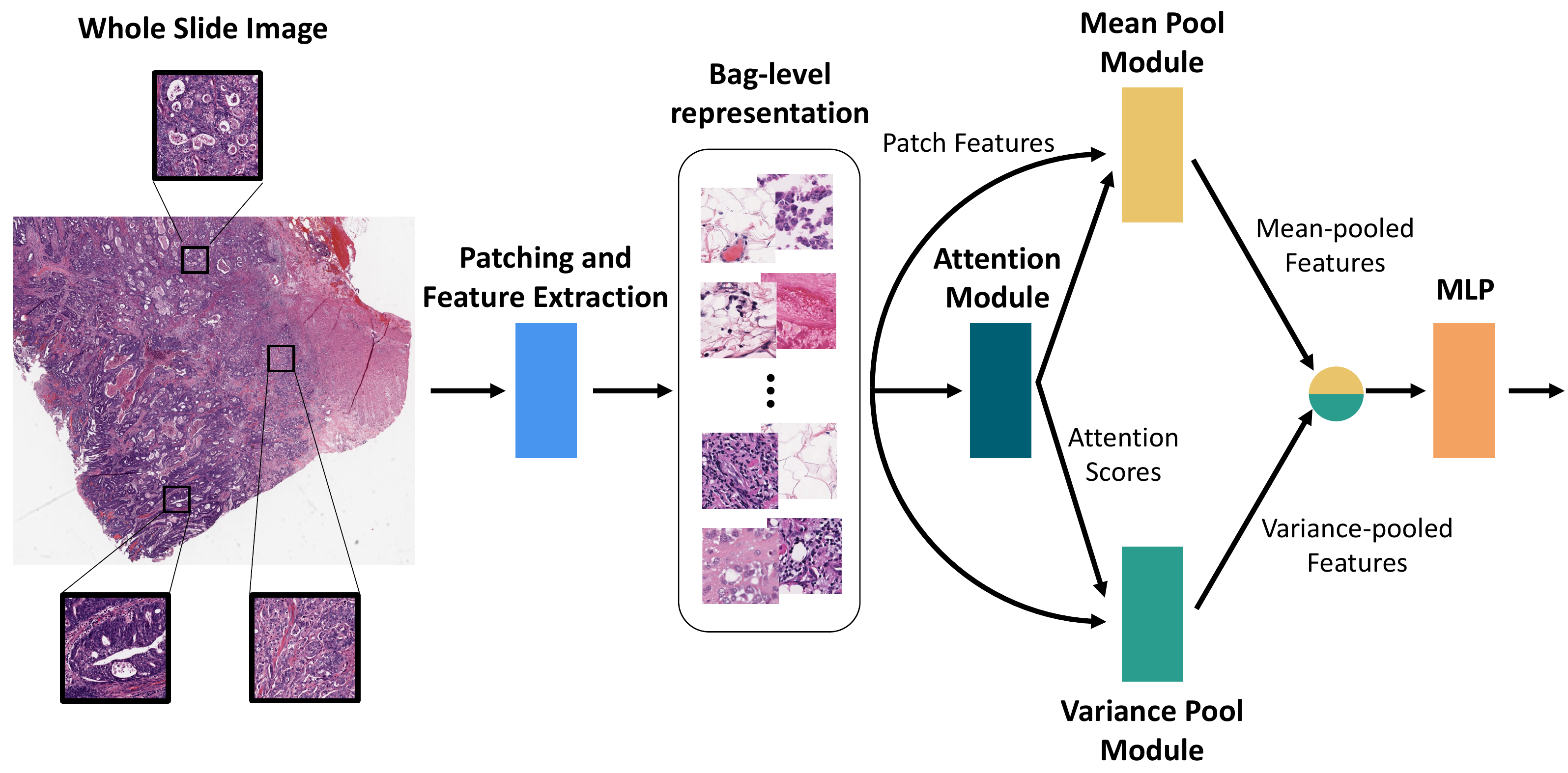}
    \caption{Variance pooling framework.
    Zoomed-in WSI patches show examples of intratumoral heterogeneity, with the colonic adenocarcinoma displaying multiple moderately- to poorly-differentiated morphologies. The mean and variance pooling branches capture different statistical characteristics of the data and are aggregated to yield the final prediction.}
    \label{fig:arch}
\end{figure}
\subsection{Attention mean pooling architecture}\label{ss:attn_mean_pool}
The attention mean pool architecture of \cite{ilse18attention} reduces a bag of instances, $\{\instance_j\}_{j=1}^n$, to a vector through a weighted mean operation.
Specifically, each instance is first passed through an (optional) encoding network, as $ h_j = \phi(\instance_j) \in \mathbb{R}^{\DimEncode}$. 
The attention neural network $\alpha$, takes in $\{h_j\}_{j=1}^n$ and computes the normalized attention weights $a = \alpha\left(\{h_j\}_{j=1}^{n}\right) \in \mathbb{R}^{n}_+$, which sum to 1.
The attention mean pool vector $p_{\text{mean}}$ is obtained as
\begin{equation} \label{eq:attn_mean_pool}
p_{\text{mean}} = \sum_{j=1}^{n} a_j  h_j. 
\end{equation}
Finally, the bag prediction (e.g. predicted survival risk) is obtained by passing $p_{\text{mean}}$ through an additional multi-layer perceptron (MLP), $z = \rho(p_{\text{mean}})$.

\subsection{Attention variance pooling}\label{ss:attn_var_pool}

We incorporate within-bag heterogeneity (intratumoral heterogeneity) by computing the variance of learned projections of the instances.
First, let us define the \textit{attention variance function}, $\AttnVarFcn{\cdot}{a}: \mathbb{R}^n \to \mathbb{R}$, of $n$ numbers,  $u \in \mathbb{R}^n$, as
\begin{equation}
\AttnVarFcn{u}{a} := \sum_{j=1}^n a_j \left(u_j - \sum_{\ell=1}^n a_{\ell} u_{\ell} \right)^2
\end{equation}
with attention weights $a \in \mathbb{R}^n_+$ summing to one.
Note, if the attention weights are all equal this is the empirical variance of the entries of $u$.

We then define a $K$-dimensional variance pool of $\{h_j\}_{j=1}^n$ as follows.
Let $v_k \in \mathbb{R}^{\DimEncode}, k \in [K] := \{1, \dots, K\},$ be variance projection vectors, which are learnable parameters in the network.
Denote the matrix of encoded instance features as $H \in \mathbb{R}^{n \times \DimEncode}$, i.e. the $j$th row of $H$ is given by $h_j$.
The $k$th entry\footnote{Without the attention weights this would be the quadratic form $v_k^T \text{cov}(H)v_k$.}
of the variance pool vector $p_{\text{var}} \in \mathbb{R}^K_+$ is obtained as
\begin{equation}\label{eq:attn_var_pool}
p_{\text{var}, k} =  \AttnVarFcn{Hv_k}{\alpha} = \sum_{j=1}^n a_j \left(h_j^Tv_k - \sum_{\ell=1}^n a_{\ell} h_{\ell}^Tv_k \right)^2.
\end{equation}
This can be viewed as dimensionality reduction on the (attention weighted) covariance matrix of $H$, i.e., from $O(\DimEncode^2)$ to $K$.

We pass $p_{\text{var}}$ through an entrywise non-linearity, $\eta(\cdot)$, such as $\sqrt{\cdot}$ or $\log(\epsilon + \cdot)$. 
This non-linearity operation is required since linear combinations of several variance pools is equivalent to just a single variance pool.
Finally, the first and second moment information of the attended $\{h_j\}_{j=1}^n$ are combined and the prediction is obtained by passing the concatenated mean and variance pool vector,
\begin{equation}\label{eq:cat_mean_var}
p_{\text{cat}} =  \left[p_{\text{mean}}; \eta(p_{\text{var}}) \right] \in \mathbb{R}^{\DimEncode + K}
\end{equation}
through the MLP,  $z = \rho(p_{\text{cat}})$\footnote{A similar approach, which we were unaware of until after publication, has been successfully used for microsatellite instability (MSI) prediction~\cite{schirris2022deepsmile}. This approach is most related to variance pooling with K=d and the variance projection coefficients fixed to be the standard basis vectors. }.

Figure~\ref{fig:arch} shows the entire architecture.
The learnable parameters are the networks $\{\phi, \alpha, \rho\}$ and $ \{v_k\}_{k=1}^K$.
Several design choices can be modified including: no attention mechanism,
separate attention mechanisms for the mean and variance pools, and different combination schemes for the mean/variance pools. 

\section{Experiments with survival prediction}\label{s:exper}

\noindent \textbf{Data:} From the public TCGA dataset, we use: Breast Invasive Carcinoma (BRCA, $N=1,061$), Glioblastoma \& Lower Grade Glioma (GBMLGG, $N=872$), Uterine Corpus Endometrial Carcinoma (UCEC, $N=504$), Colon Adenocarcinoma \& Rectum Adenocarcinoma (COADREAD, $N=612$), and Bladder Urothelial Carcinoma (BLCA, $N=386$).
The choice of these types was based on the number of patients.
We use \textit{progression-free interval} (PFI) \citep{liu2018integrated}.

We use the concordance index (c-Index) to evaluate survival prediction models.
The c-Index, which lies between 0 and 1 (larger is better), measures the agreement between predicted risk scores and actual survival outcomes using only order information (see discussion around \eqref{eq:rank_loss} below.)
We split each dataset into a $70\%/30\%$ train/test split.
To ensure that our results are not sensitive to a particular data split, we randomly split the data into train/test folds 10 times. \\

\noindent \textbf{Implementation Details:} We extract $d=1,024$ features from each WSI image patch ($256\times 256$ pixels at $20\times$ magnification) using an ImageNet-pretrained ResNet-50 CNN encoder, truncated after the third residual block and spatially averaged. 
We use a batch size of $B=32$ patients\footnote{Note the patient batch size is not equal to the number of summands in \eqref{eq:rank_loss}; there are between 0 and ${B \choose 2}$ summands depending on the number of comparable pairs.}.
To fit each mini-batch into GPU memory during training we cap--by random subsampling-- the number of instances in each bag.
For each cancer type, the maximum number of instances per bag is set to be the $75\%$ quantile of the overall bag sizes (WSIs typically have $15-20,000$ patches). In the Supplemental information, we report the summary statistics for number of instances per WSI slide in the TCGA dataset.
All patches in the bag are used at test time.
We use the Adam optimizer with a learning rate of $2\times 10^{-4}$, weight decay of $1 \times 10^{-5}$, and train for 30 epochs.

We use three MIL architectures: Deep Sets~\citep{zaheer17deep}, Attention MeanPool~\citep{ilse18attention}, and DeepGraphConv~\citep{zhao2020predicting}. 
For the Deep Sets framework, the attention weights are not learned and instead set to identical values of $1/n$.
For the DeepGraphConv framework, the bag of features is first passed through two layers of graph convolutional neural network layers, the output of which are passed through the attention mean/variance pooling modules.
These are some of the most popular MIL   architectures used in computational pathology and thus provide great testbeds for variance pooling module extension.

We use $K=10$ variance pooling projections and an $\eta(\cdot) = \log(0.01 + \cdot)$ nonlinearity. 
We performed an internal ablation study for the nonlinearity comparing the logarithm, square root, and sigmoid functions, but did not find any significant difference in performance.
We initialize the projection vectors $v_k \sim \mathcal{N}(0, 1/\DimEncode)$ to make them approximately orthonormal with high probability. \\

\noindent \textbf{Loss function:} Suppose the bag prediction network, $f(\cdot)$, outputs a risk score for each patient.
We train the network with the survival ranking loss \citep{steck2007ranking,luck2018learning}, which is a continuous approximation of the negative c-Index given by
\begin{equation}\label{eq:rank_loss}
    \mathcal{L}\left(\{\bag\ssp{i}, t\ssp{i}, c\ssp{i}\}_{i=1}^N \right) = -\frac{1}{|\mathcal{C}|} \sum_{(w, b)\in\mathcal{C}} \psi \left(f(\bag\ssp{w}) - f(\bag\ssp{b}) \right),
\end{equation}
where $\mathcal{C} = \left\{(w, b) | c\ssp{w}=0 \text{ and } t\ssp{w} < t\ssp{b}, \text{ for } w, b \in [N] \right\}$ is the set of \textit{comparable pairs} i.e. pairs of patients where we know one patient has worse survival than the other.
If $\psi(\cdot) = \mathbf{1}(\cdot > 0)$ then \eqref{eq:rank_loss} would be exactly the negative c-Index.
We set $\psi(\cdot)$ to be a sigmoid, which is a differentiable approximation of this indicator.

We chose the ranking loss because it is separable, which allows the use of proper mini-batches unlike the Cox loss.
Additionally, the c-Index is a natural objective for survival analysis.
An alternative is the negative log-likelihood of a discrete time-to-event model \citep{zadeh2020bias}.
This loss function
coarsely bins the survival times and does not fully utilize the available survival ordering information.
\begin{table}
\caption{Test set c-Index performance (mean $\pm$ std) across 10 re-sampled data splits on the 100 point scale.}\label{tab:results}
\centering
\begin{tabular}{p{0.35\textwidth}>{\centering}p{0.15\textwidth}>{\centering}p{0.2\textwidth}>{\centering\arraybackslash}p{0.2\textwidth}}
\hline
Frameworks & BRCA & GBMLGG & COADREAD\\
\hline
Deep Sets & $59.5 \pm 3.1$ & $72.8 \pm 1.0$ & $50.8 \pm 4.1$\\
Deep Sets w/ VarPool & $\mathbf{60.0 \pm 3.8}$ & $\mathbf{73.8 \pm 0.9}$ & $\mathbf{56.1 \pm 3.0}$ \\
\hline
Attn MeanPool & $59.2 \pm 3.7$ & $73.0 \pm 1.2$ & $52.0 \pm 2.3$ \\
Attn MeanPool w/ VarPool & $\mathbf{60.4 \pm 3.9}$ & $\mathbf{73.6 \pm 1.0}$ & $\mathbf{56.3 \pm 2.3}$ \\
\hline
DeepGraphConv & $57.9 \pm 4.2$& $70.7 \pm 2.3$& $51.4 \pm 3.0$\\
DeepGraphConv w/ VarPool& $\mathbf{58.0 \pm 5.2}$& $\mathbf{70.9 \pm 2.3}$ & $\mathbf{51.6 \pm 3.2}$\\
\hline
& BLCA & UCEC & Overall \\
\hline
Deep Sets & $56.6\pm 5.3$& $51.3 \pm 4.3$ & $58.2$\\
Deep Sets w/ VarPool & $\mathbf{57.5 \pm 5.1}$ & $\mathbf{54.1 \pm 3.0}$ & $\mathbf{60.3}$\\
\hline
Attn MeanPool & $57.2 \pm 5.3$ & $50.7 \pm 3.9$ & $58.4$\\
Attn MeanPool w/ VarPool & $\mathbf{57.8 \pm 4.6}$ & $\mathbf{52.5 \pm 3.7}$ & $\mathbf{60.1}$\\
\hline
DeepGraphConv & $52.1 \pm 7.7 $& $54.2 \pm 2.1$ & $57.3$\\
DeepGraphConv w/ VarPool& $\mathbf{53.1 \pm 7.6}$& $\mathbf{55.0 \pm 2.0}$& $\mathbf{57.7}$\\
\hline
\end{tabular}
\end{table}

Table~\ref{tab:results} shows the results.
We observe that for every cancer type and all models, the variance pooling versions outperform their counterparts, with up to $10.4 \%$ relative increase (Deep Sets for COADREAD). 
We conjecture that the smaller performance gain in DeepGraphConv comes from ``feature smoothing" in GCNs, which reduces the heterogeneity in patch features.
In the Supplemental information, we present an ablation study with the Cox loss and show that similar improvements are observed with the addition of variance pooling module.

The fact that explicitly incorporating intratumoral heterogeneity gives the greatest benefit in COADREAD and UCEC is interesting, since subsets of these tumor types have very high mutation rates due to microsatellite instability, a molecular change with impacts on survival ~\citep{levine2013integrated,guinney2015consensus}.
This might lead to the development of multiple subclones of tumor cells that have different morphologies, as the effects of those mutations propagate from DNA to protein.

\section{Interpretability and biological insights}\label{s:interpret}

This section presents two interpretability approaches to probe what biological signals the trained models learn.
We focus on the Attention MeanPool and VarPool architectures, though similar approaches can be used for other architectures.\\

\noindent\textbf{Interpretability via attention scores}
For each patient, we visualize the patches with the largest attention scores, i.e., the $a$ in \eqref{eq:attn_mean_pool} and \eqref{eq:attn_var_pool}, as depicted in Figure \ref{fig:top_attn}.
Note that this interpretability approach is not possible for the Deep Sets framework, since the attention weights are set to be uniform.

\noindent\textbf{Interpretability via variance projection scores} For variance pooling, we can examine signals captured by each variance pool projection.
Each projection can be viewed as sorting the patches of a WSI, the biological signal of which we aim to uncover.
We project the patches onto a given projection vector and visualize the patches along the spectrum.
A similar interpretability approach was taken in \cite{carmichael2021joint}.
Following \eqref{eq:attn_var_pool}, for a projection vector $v$ and $j$th instance, we define the \textit{signed attention weighted squared residuals} (SAsqR) as
\begin{equation*}
\operatorname{}{SAsqR}_j = \text{sign}(r_j) a_j r_j^2, \text{ where } r_j := h_j^Tv - \sum_{\ell=1}^n a_{\ell} h_{\ell}^Tv.
\end{equation*}
These values can be used to sort the patches e.g. to identify the patches with the most extreme negative/positive $\text{SAsqR}$ values.

\subsection{Interpretability visualizations}\label{ss:interpret_results}

We illustrate the two interpretability tools discussed in the previous section.
Our analysis outputs a large number of images based on these tools.
The figures below illustrate the trends that were reviewed by two pathologists.

\begin{figure}[!ht]
 \centering
\begin{subfigure}[t]{0.35\textwidth}
\centering
\includegraphics[width=\linewidth, height=0.57\linewidth]{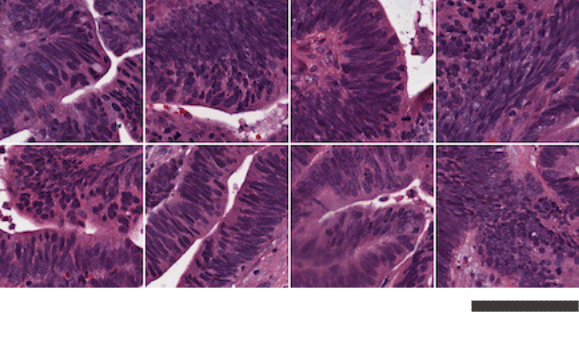}
\caption{
Top MeanPool patches.
}
\label{fig:TCGA-A6-5660-amil_nn}
\end{subfigure}
\hspace{2em}
\begin{subfigure}[t]{0.35\textwidth}
\centering
\includegraphics[width=\linewidth, height=0.57\linewidth]{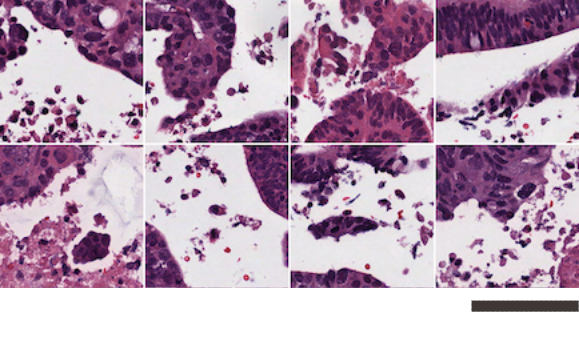}
\caption{
Top VarPool patches.
}
\label{fig:TCGA-A6-5660-amil_var_nn}
\end{subfigure}
\caption{
The highest attended patches for the attention MeanPool and VarPool for a single COADREAD patient.
Here, MeanPool typically attends to solid tumor while VarPool attends to fragmented, more poorly-differentiated tumor. The scale bar represents $100\, \mu m$.
}
\label{fig:top_attn}
\end{figure}

Figure \ref{fig:top_attn} compares the top attended patches for MeanPool and VarPool in one COADREAD patient.
We observed that MeanPool attends to more compact, dense, and intact tumor with surrounding stroma, while VarPool attends to fragmented tumor, often with surrounding extracellular mucin and debris. 

\begin{figure}[!ht]
 \centering
\begin{subfigure}[t]{\textwidth}
\centering
\includegraphics[width=\linewidth, height=0.17\linewidth]{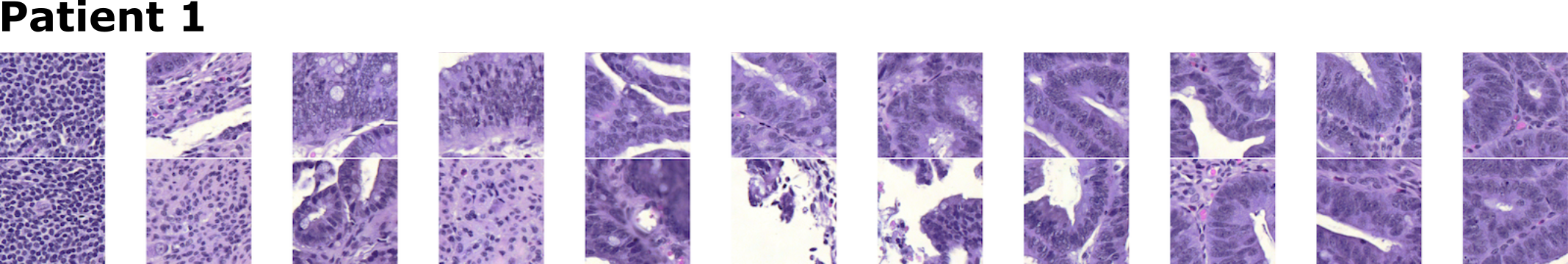}
\label{fig:TCGA-AA-3818-rows}
\end{subfigure}
\hfill
\begin{subfigure}[t]{\textwidth}
\centering
\includegraphics[width=\linewidth, height=0.23\linewidth]{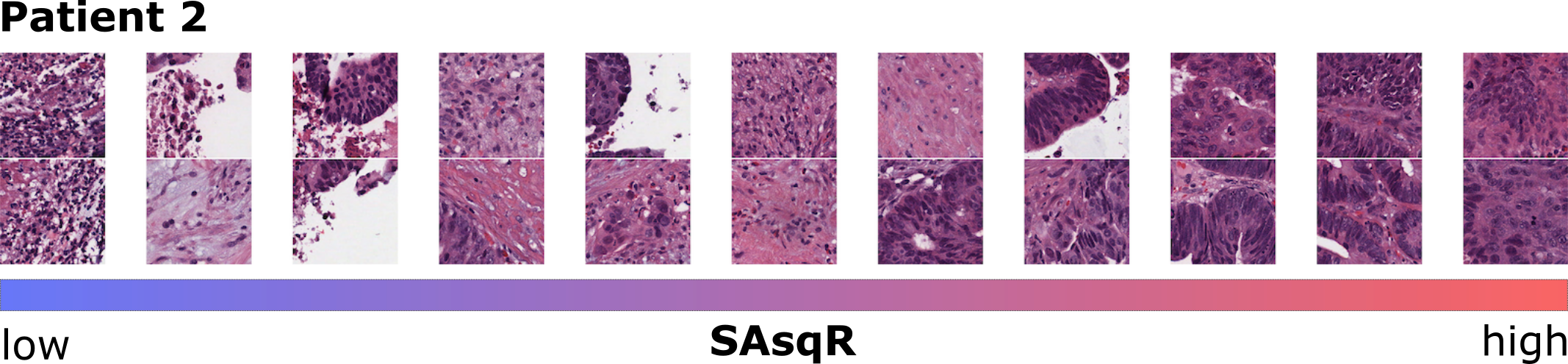}
\label{fig:TCGA-A6-5660-rows}
\end{subfigure}
\caption{
Patches along one variance projection for two COADREAD patients.
The patches are ordered from the most negative (left) SAsqR scores to most positive (right). 
The leftmost patches contain dense inflammatory infiltrate comprised mostly of lymphocytes, while the rightmost patches show denser tumor regions.
}
\label{fig:var_proj_coadread_proj_6}
\end{figure}

\begin{figure}[!ht]
 \centering
\begin{subfigure}[t]{\textwidth}
\centering
\includegraphics[width=\linewidth, height=0.17\linewidth]{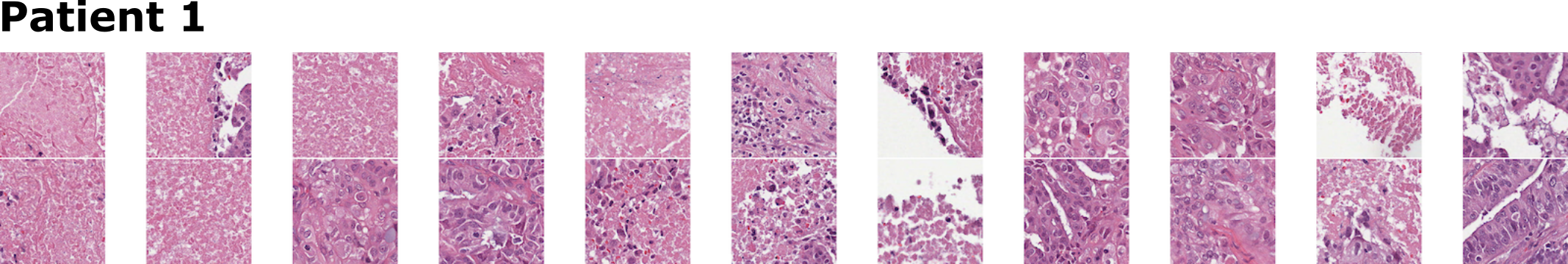}
\label{fig:TCGA-AJ-A3BG-rows}
\end{subfigure}
\hfill
\begin{subfigure}[t]{\textwidth}
\centering
\includegraphics[width=\linewidth, height=0.23\linewidth]{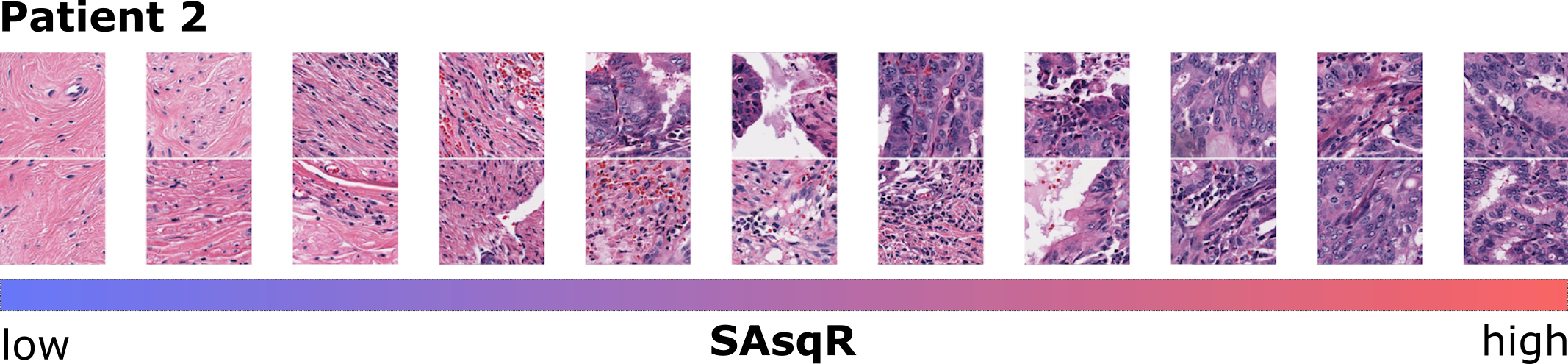}
\label{fig:TCGA-B5-A5OC-rows}
\end{subfigure}
\caption{
Patches along one variance projection for two UCEC patients.
This projection shows two patterns that are illustrated by these patients.
For the first patient, the leftmost patches display necrosis and the right most patches display tumor.
For the second patient, the leftmost patches display predominantly muscle and connective tissue and the right most patches display tumor.
}
\label{fig:var_proj_ucec_proj_0}
\end{figure}

Figure \ref{fig:var_proj_coadread_proj_6} shows patches along a single variance projection for two COADREAD patients.
Figure \ref{fig:var_proj_ucec_proj_0} is similar for UCEC.
We have sorted the patches by their SQsqR values and shown several patches for 11 quantiles: 0, 10, ..., 100.
We might expect the rightmost and leftmost patches to represent visual contrast.
We observe dense tumor region on one end of the spectrum and other micro-environment features, such as lymphocytes, muscle, or necrosis on the other end.

\section{Conclusion} \label{s:conclusion}
We introduce a novel variance pooling module that can be incorporated into existing multiple instance learning (MIL) frameworks, to encode heterogeneity information through the second order statistics of low rank projections. 
We show that variance pooling can improve survival prediction across several cancer types.
We also provide interpretability tools based on attention and projection scores to understand the heterogeneous biological signals captured by our framework.
We hope this study will lead to further investigations of intratumoral heterogeneity -- and other tumor microenvironment features -- for computational pathology.

\section*{Acknowledgement} \label{s:ack}
We thank Katherine Hoadley for helpful suggestions.
This work was supported in part by internal funds from BWH Pathology, NIGMS R35GM138216 (F.M.), BWH President’s Fund, MGH Pathology, BWH Precision Medicine Program, Google Cloud Research Grant, Nvidia GPU Grant Program and funding from the Fredrick National Lab. R.C. was additionally supported by the NSF graduate research fellowship. T.Y.C. was additionally funded by the NIH National Cancer Institute (NCI) Ruth L. Kirschstein National Service Award, T32CA251062. The content is solely the responsibility of the authors and does not reflect the official views of the NIH, NIGMS, NCI, or NSF.

\clearpage
\bibliographystyle{plainnat}
\bibliography{varpool}

\clearpage

\section*{Appendix}

\begin{figure}[H]
 \centering
\begin{subfigure}[t]{0.35\textwidth}
\centering
\includegraphics[width=\linewidth, height=0.5\linewidth]{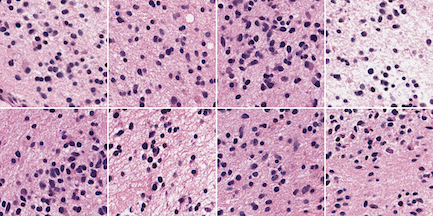}
\caption{
Top MeanPool patches.
}
\label{fig:TCGA-DB-A64X-mean_pool-top_attn-patches}
\end{subfigure}
\hspace{2em}
\begin{subfigure}[t]{0.35\textwidth}
\centering
\includegraphics[width=\linewidth, height=0.5\linewidth]{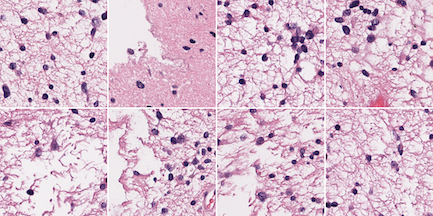}
\caption{
Top VarPool patches.
}
\label{fig:TCGA-DB-A64X-var_pool-top_attn-patches}
\end{subfigure}
\caption{
The highest attended patches for the attention MeanPool and VarPool for a single GBMLGG patient.
MeanPool attends to patches with high tumor cellularity, while VarPool attends to patches with lower cellularity.
}
\end{figure}

\begin{figure}[H]
 \centering
\begin{subfigure}[t]{\textwidth}
\centering
\includegraphics[width=\linewidth, height=0.14\linewidth]{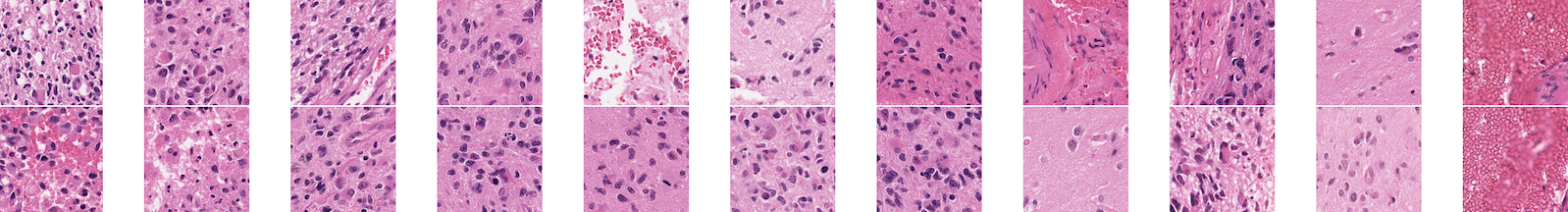}
\end{subfigure}
\hfill
\begin{subfigure}[t]{\textwidth}
\centering
\includegraphics[width=\linewidth, height=0.14\linewidth]{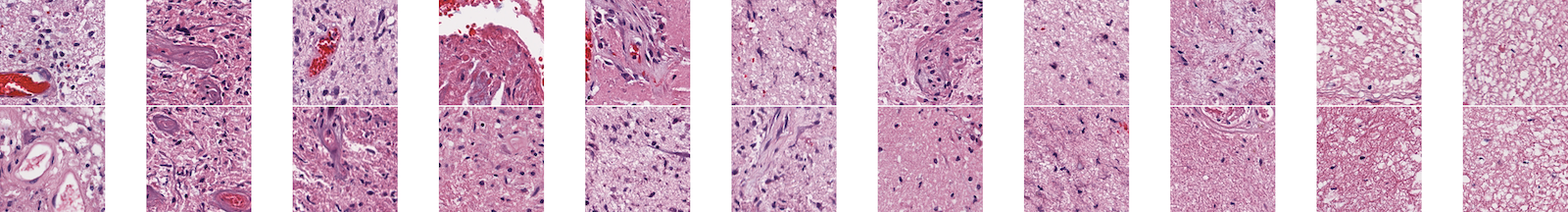}
\end{subfigure}
\caption{
Patches along one variance projection for two GBMLGG patients. As in the main body, the patches are ordered from the most negative (left) SAsqR scores to most positive (right).
The top patient is the patient predicted to have a high risk score, while the bottom patient is predicted to have a low risk score.
This projection sorts the patches from lower cellularity (right) to higher cellularity (left) -- recall we focus on the right/left extremes.
}
\label{fig:var_proj_gbmlgg_proj_6}
\end{figure}

\begin{table}
\caption{Loss function ablation study. Test set c-Index performance (mean $\pm$ std) on the 100 point scale for \textbf{cox loss}, where the training details were the same as the rank loss experiment. This shows that the variance pooling mechanism is loss-agnostic, equally showing improvement for both rank and cox loss.}
\centering
\begin{tabular}{p{0.35\textwidth}>{\centering}p{0.15\textwidth}>{\centering}p{0.2\textwidth}>{\centering\arraybackslash}p{0.2\textwidth}}
\hline
Frameworks & BRCA & GBMLGG & COADREAD\\
\hline
Deep Sets &  $60.34\pm 4.45$& $72.76 \pm 1.13$ & $54.01 \pm 2.6$ \\
Deep Sets w/ VarPool &$\mathbf{60.96\pm5.41}$ &$\mathbf{73.88\pm 1.26}$ & $\mathbf{57.95 \pm 3.01}$\\
\hline
Attn MeanPool & $59.89\pm 4.51$ & $\mathbf{73.25\pm 1.16}$ & $54.71\pm 2.55$ \\
Attn MeanPool w/ VarPool & $\mathbf{61.1\pm 4.48}$ &$73.21\pm1.18$& $\mathbf{58.49 \pm 3.12}$ \\
\hline
DeepGraphConv & $57.62 \pm 4.95$ & $\mathbf{71.26 \pm 2.3}$ & $56.03 \pm 2.59$ \\
DeepGraphConv w/ VarPool & $\mathbf{59.78 \pm 5.2}$ & $ 71.15 \pm 2.34$ & $\mathbf{56.23 \pm 3.3}$\\
\hline
& BLCA & UCEC & Overall \\
\hline
Deep Sets & $56.82\pm 4.31$ & $52.05\pm 4.35$& $59.20$\\
Deep Sets w/ VarPool & $\mathbf{57.59\pm 3.84}$& $\mathbf{55.14\pm 3.43}$& $\mathbf{61.10}$ \\
\hline
Attn MeanPool & $56.95\pm 4.26$& $51.16\pm 2.85$ & $59.19$\\
Attn MeanPool w/ VarPool & $\mathbf{57.47\pm 3.43}$& $\mathbf{53.53\pm 3.49}$& $\mathbf{60.76}$ \\
\hline
DeepGraphConv & $52.35 \pm 7.1$ & $51.15 \pm 1.89$ & $57.68$\\
DeepGraphConv w/ VarPool& $\mathbf{52.52 \pm 5.96}$ & $\mathbf{53.2 \pm 2.33}$ & $\mathbf{58.58}$\\
\hline
\end{tabular}
\end{table}

\begin{table}
\caption{The summary statistics for number of instances per WSI slide in the TCGA dataset. 
For training, we restricted the number of instances per slide to the $75 \% $ quantile for each cancer type. 
To fit a batch in one tensor, zero-padding and manually zeroing the attention was required for slides with fewer instances. 
For slides with more instances, we performed random subsampling. 
These measures were necessary to perform training with minibatches of slides (instead of batch size of 1 slide, as typically done in the literature). }
\centering
\begin{tabular}{p{0.2\textwidth}>{\centering}p{0.15\textwidth}>{\centering}p{0.15\textwidth}>{\centering}p{0.15\textwidth}>{\centering}p{0.15\textwidth}>{\centering\arraybackslash}p{0.15\textwidth}}
\hline
& BRCA & GBMLGG & COADREAD & BLCA & UCEC\\
\hline
mean     & $10,826.28$& $7,448.28$ & $8.575.54$ & $16,599.11$ & $16,003.92$  \\
median     & $10,388$ & $4,661$ & $5,535$ & $16,742$ & $16,856$ \\
$75 \%$ quantile     & $14,871$ & $11,803$ & $14,076$ & $20,766$ & $21,624$ \\
min     & $260$ & $23$ & $174$ & $491$ & $458$ \\
max     & $67,268$ & $36,996$ & $36,495$ & $39,793$ & $45,482$ \\
std & $6,999.84$ & $6,847.36$ & $6,799.17$ & $7.118.88$ & $8,219.33$\\
\hline
\end{tabular}
\end{table}

\end{document}